\documentclass[3p,times,twocolumn]{elsarticle}

\usepackage{ecrc}

\volume{00}

\firstpage{1}

\journalname{Nuclear Physics B Proceedings Supplement}

\runauth{}

\jid{nuphbp}

\jnltitlelogo{Nuclear Physics B Proceedings Supplement}

\usepackage{amssymb}
\usepackage[figuresright]{rotating}

\begin{document}

\begin{frontmatter}



\dochead{}

\title{$\alpha_s$ from $\tau$ decays revisited}


\author[aachen]{D.R.~Boito}
\author[munich]{O.~Cat\`a}
\author[sfsu]{M.~Golterman}
\author[icrea]{M.~Jamin\corref{cor}}
\author[york,cssm]{K.~Maltman}
\author[sfsu]{J.~Osborne}
\author[sfsu,uab]{S.~Peris}

\address[aachen]{Institut f\"ur Theoretische Teilchenphysik und Kosmologie,
RWTH Aachen University, D-52074 Aachen, Germany}
\address[munich]{Ludwig-Maximilians-Universit\"at M\"unchen, Fakult\"at f\"ur
Physik, Arnold Sommerfeld Centre for Theoretical Physics, D-80333 M\"unchen,
Germany}
\address[sfsu]{Department of Physics and Astronomy, San Francisco State
University, San Francisco, CA 94132, USA}
\address[icrea]{ICREA and IFAE, Universitat Aut\`onoma de Barcelona, E-08193
Bellaterra, Barcelona, Spain}
\address[york]{Department of Math and Statistics, York University, Toronto,
ON Canada M3J 1P3}
\address[cssm]{CSSM, University of Adelaide, Adelaide, SA Australia 5005}
\address[uab]{Departament de F\'isica, Universitat Aut\`onoma de Barcelona,
E-08193 Bellaterra, Barcelona, Spain}

\cortext[cor]{Corresponding author}

\begin{abstract}
Being a determination at low energies, the analysis of hadronic $\tau$ decay
data provides a rather precise determination of the strong coupling $\alpha_s$
after evolving the result to $M_Z$. At such a level of precision, even small
non-perturbative effects become relevant for the central value and error. While
those effects had been taken into account in the framework of the operator
product expansion, contributions going beyond it, so-called duality violations,
have previously been neglected. The following investigation fills this gap
through a finite-energy sum rule analysis of $\tau$ decay spectra from the OPAL
experiment, including duality violations and performing a consistent fit of
all appearing QCD parameters. The resulting values for $\alpha_s(M_\tau)$ are
$0.307(19)$ in fixed-order perturbation theory and $0.322(26)$ in
contour-improved perturbation theory, which translates to the $n_f=5$ values
$0.1169(25)$ and $0.1187(32)$ at $M_Z$, respectively.
\end{abstract}


\end{frontmatter}

% \linenumbers


\section{Introduction}
\label{sect1}

The presented contribution summarises the recent article \cite{bcgjmop11},
which laid out a new framework to determine the strong coupling $\alpha_s$
from hadronic $\tau$ decays. After the determination of the order $\alpha_s^4$
coefficient to the QCD Adler function \cite{bck08}, interest in the extraction
of $\alpha_s$ from $\tau$ decays was revived \cite{bck08,ddhmz08,bj08,my08}.
However, the results obtained in two approaches to resumming the perturbative
series, fixed-order perturbation theory (FOPT) and contour-improved
perturbation theory (CIPT), are only barely compatible. 

While this issue will be left untouched here, the $\alpha_s$ determination
from $\tau$ decays also requires numerical inputs for the non-perturbative
contributions that arise in the framework of the operator product expansion
(OPE), the QCD condensates. Even though the OPE terms are usually included
in the analysis, contributions which go beyond it, so-called {\em duality
violations} (DVs), and which arise due to the presence of resonances on the
physical, Minkowskian axis, have so far been neglected. However, for a
consistent estimate of the OPE contributions, effects of DVs should be included
since typically several spectral integrals of the experimental data with
differing weight functions, and possibly at varying energies, are employed,
to which the DVs contribute differently.\footnote{See also the contribution by
K.~Maltman to these proceedings \cite{kim11}.} An effort to fill this gap was initiated in
the work of ref.~\cite{bcgjmop11}.

\section{Theoretical framework}
\label{sect2}

The central observables for the analysis of hadronic $\tau$ decay are the
total decay rate into hadrons,
\begin{equation}
\label{Rtau}
R_\tau \,\equiv\, \frac{\Gamma [ \tau^- \to \nu_\tau\, \rm{hadrons} ]}
{\Gamma [ \tau^- \to \nu_\tau \, e^- \, \overline{\nu}_e ]} \,,
\end{equation}
and corresponding differential decay spectra. Experimentally, those spectra
can be decomposed into the non-strange vector ($V$) and axial-vector ($A$), as
well as the strange components.

The theoretical description is based on the two-point correlation functions
\begin{eqnarray}
\label{Pimunu}
\Pi_{V,A}^{\mu\nu}(q) \!\!\!&\equiv&\!\!\! i\! \int \!{\rm d}^4\!x \,e^{iqx}
\,\langle 0| T \{ J_{V,A}^\mu(x) J_{V,A}^{\nu\,\dagger}(0)\}|0 \rangle \\
\!\!\!&=&\!\!\! ( q^\mu q^\nu - g^{\mu \nu} q^2 )\,\Pi_{V,A}^{(1)}(q^2) +
q^\mu q^\nu \Pi_{V,A}^{(0)}(q^2) \,, \nonumber
\end{eqnarray}
where the relevant $V$ and $A$ currents are given by $J_V^\mu(x)=\overline{u}(x)
\gamma^\mu d(x)$ and $J_A^\mu(x)=\overline{u}(x) \gamma^\mu \gamma_5 d(x)$, and
the superscripts $(1,0)$ label the spin.

Employing the physical spectral functions $\rho_{V,A}^{(J)}(s)\equiv
{\rm Im}\, \Pi_{V,A}^{(J)}(s)/\pi$, the $V$ and $A$ contributions to
eq.~(\ref{Rtau}) can be expressed as the $s_0=M_\tau^2$ version of the
weighted integral \cite{bnp92}
\begin{eqnarray}
\label{RVAs0}
R_{V,A}(s_0) \!\!\!&=&\!\!\! 12 \pi^2\,S_{\rm EW}\,|V_{ud}|^2
\int\limits_0^{s_0} \frac{{\rm d}s}{s_0} \;\Big(1 - \frac{s}{s_0}\Big)^2
\times \nonumber \\
&& \hspace{-6mm} \Big[\,\Big( 1 + 2\,\frac{s}{s_0} \Big)\,\rho_{V,A}^{(1+0)}(s)
- 2\,\frac{s}{s_0}\,\rho_{V,A}^{(0)}(s)\,\Big] \,.
\end{eqnarray}
The scalar/pseudoscalar ($J=0$) terms are suppressed by factors of $m_{u,d}^2$,
and hence negligible, apart from the pion-pole contribution. Thus, the
experimental decay distributions essentially determine $\rho_{V,A}^{(1+0)}(s)$.

Making use of Cauchy's theorem, the analytic structure of $\Pi_{V,A}^{(1+0)}(s)$
implies that for any analytic weight $w(s)$ the following finite energy sum
rule (FESR) is satisfied \cite{sha77}:
\begin{equation}
\label{FESR}
\int\limits_0^{s_0}\! {\rm d}s\, w(s)\,\rho_{V,A}^{(1+0)}(s) \,=\,
\frac{i}{2\pi}\!\oint\limits_{|s|=s_0}\!{\rm d}s\,w(s)\,\Pi_{V,A}^{(1+0)}(s)\,.
\end{equation}
The central idea then is to evaluate the LHS of (\ref{FESR}) from experimental
data, and the RHS within the theoretical QCD description, thereby allowing to
determine fundamental QCD parameters like $\alpha_s$.

Conventionally, the theoretical side is calculated in the framework of the
operator product expansion. However, the contour integral has to be performed
down to the physical, Minkowskian axis, where the OPE breaks down due to the
presence of resonances, and quark-hadron duality violations arise. Even though
the kinematic weight in (\ref{RVAs0}) has a double zero at $s=s_0$, which
may suppress DVs, for a general weight $w(s)$ those contributions should be
included.

The theoretical 2-point correlators $\Pi_{V,A}^{(1+0)}(s)$ will thus be
represented as
\begin{equation}
\label{PiTH}
\Pi_{V,A}^{(1+0)}(s) \,=\, \Pi_{V,A}^{\rm OPE}(s) + \Delta_{V,A}(s) \,,
\end{equation}
where the second term corresponds to the DV contribution. Further assuming
that it vanishes sufficiently fast at infinity, the required contour integral
containing $\Delta_{V,A}(s)$ can be expressed as
\begin{equation}
\label{Delta}
\frac{i}{2\pi}\!\oint\limits_{|s|=s_0}\!{\rm d}s\,w(s)\,\Delta_{V,A}(s) \,=\,
-\int\limits_{s_0}^\infty\! {\rm d}s\, w(s)\,\rho_{V,A}^{\rm DV}(s) \,.
\end{equation}
Guided by a model for the light-quark V/A correlators \cite{bsz98}, being based
on Regge theory and large-$N_c$, we employ for $\rho_{V,A}^{\rm DV}(s)$ the
{\em Ansatz} \cite{cgp05,cgp08}
\begin{equation}
\label{rhoDV}
\rho_{V,A}^{\rm DV}(s) \,=\, \kappa_{V,A}\, {\rm e}^{-\,\gamma_{V,A} s}
\sin\left(\alpha_{V,A} + \beta_{V,A} s\right) \,.
\end{equation}
Such a model for DVs is also supported by a study of the Coulomb system
\cite{jam11}. Besides the QCD parameters, per channel this adds four additional
parameters into the theoretical description, all of which should be extracted
through fits to experiment.

\section{Numerical analysis}
\label{sect3}

The central strategy to extract all parameters, the strong coupling $\alpha_s$,
the OPE parameters, as well as DVs, is a simultaneous fit to weighted spectral
integrals (moments) of the form (\ref{FESR}) with in general different
weights $w(s)$ and/or different $s_0$, with several criteria restricting the
set of sensible choices. 

First of all, for simplicity, only weights polynomial in $s$ will be considered.
Next, as we intend to extract the parameters for DVs, a weight without
pinch-suppression (a zero at $s=s_0$) should be included. The simplest choice
here is $w(s)=1$. The order of the selected polynomial also determines the
dimension up to which OPE terms contribute at leading-order (neglecting
logarithmically suppressed effects). For a polynomial of highest order $s^n$,
operators up to dimension $D=2n+2$ contribute. As very little is known about
OPE contributions beyond $D=8$, $w(s)$ will be restricted to at most 3rd order.
The kinematical weight of eq.~(\ref{RVAs0}) is of this type. Furthermore,
below some minimal $s_0\approx 1.5\,{\rm GeV}^2$, both, the OPE as well as the
perturbative expansion become questionable and also the DV model is no longer
adequate. Hence only moments down to such a value should be included. Finally,
the moments for considered weights/$s_0$ combinations are correlated, and for
the fits to work these correlations should not become too strong. This is the
most delicate point about our analysis.

The required $V$ and $A$ spectral functions have been experimentally determined
by the ALEPH \cite{aleph98,aleph05} and OPAL \cite{opal99} collaborations at
LEP. However, the most recent publicly available ALEPH data \cite{aleph05} do
not fully include correlations due to unfolding. Therefore, we chose to
perform our analysis on the basis of the OPAL data \cite{opal99} only.

\begin{figure}[!ht]
\begin{center}
\includegraphics[width=0.9\columnwidth,angle=0]
{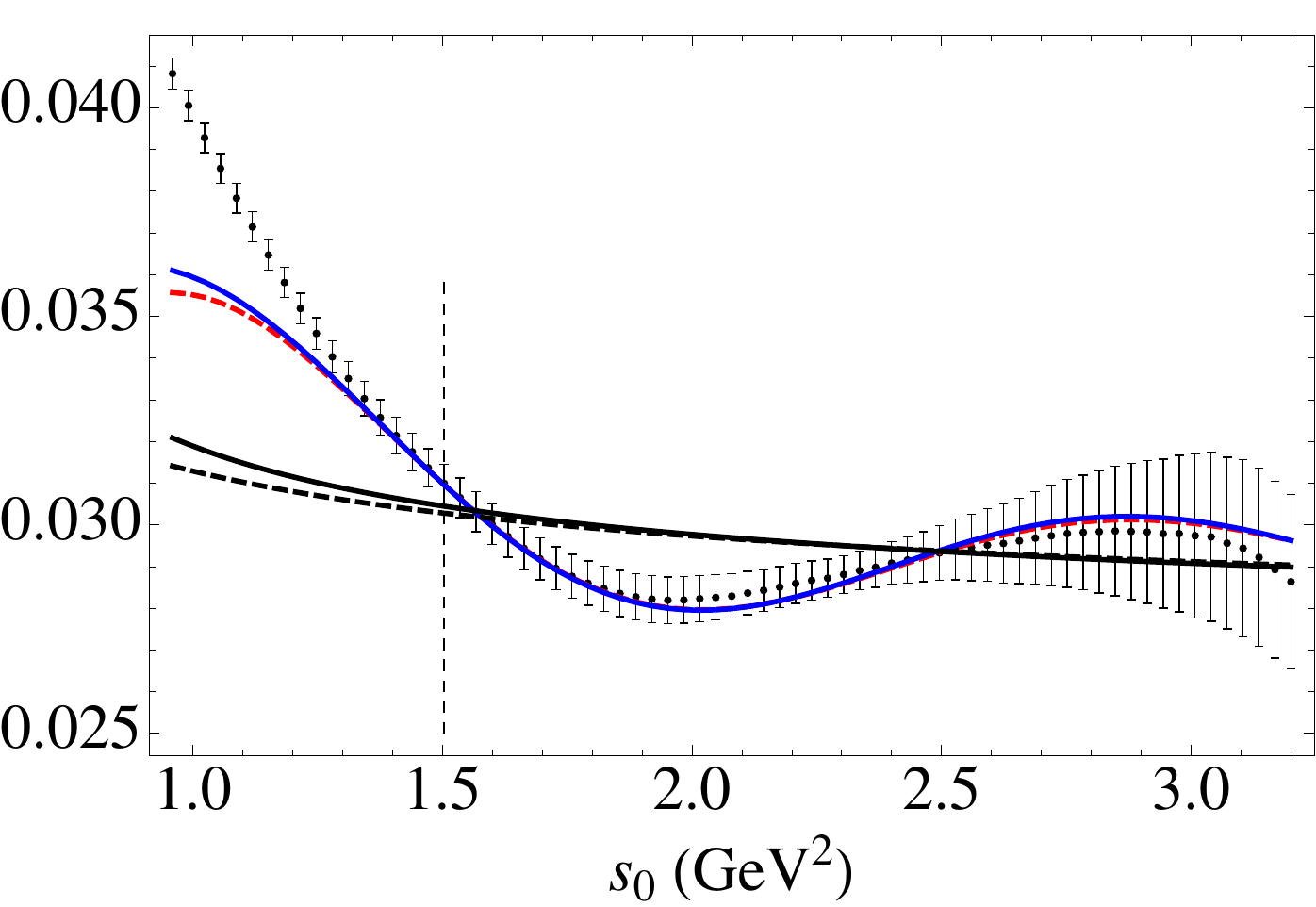}
\vspace{-0.2cm}
\caption{Comparison of experimental and theoretical $w(s)=1$ $V$ moments. FOPT
fits are shown in blue (solid) and CIPT fits in red (dashed). The (much flatter)
black curves display only the OPE parts of the FOPT (solid) and CIPT (dashed)
fit results.\label{fig1V}}
\end{center}
\end{figure}

Let us begin by discussing the most basic fit employing only $w(s)=1$ as the
weight. In this case, $s_0$'s corresponding to the upper endpoints of  all
experimental bins down to $s_{\rm min}\approx 1.5\,{\rm GeV}^2$ can be included
if the fit is performed only on the $V$ spectrum, and the fit parameters then
only consist of $\alpha_s$ and the four $V$-channel DV parameters. The
comparison of the moments as computed from the OPAL data and theoretical
prediction with fitted parameters is displayed in figure~1. The blue (solid)
and red (dashed) lines correspond to FO and CI perturbative results
respectively. For comparison the flatter black solid and dashed lines
correspond to the same fits omitting DVs, which clearly demonstrates that it
is necessary to include them in order to obtain a reasonable fit.
The $\alpha_s$ values resulting from the fit read
\begin{eqnarray}
\label{asw0}
\alpha_s(M_\tau)\!\!\!&=&\!\!\! 0.307\pm 0.018\pm 0.004\quad\mbox{(FOPT)}\,,\\
\alpha_s(M_\tau)\!\!\!&=&\!\!\! 0.322\pm 0.025\pm 0.004\quad\mbox{(CIPT)}\,,
\end{eqnarray}
where the first error is the fit uncertainty and the second results from a
variation of $s_{\rm min}$. The corresponding DV parameters can be found in
table~1 of \cite{bcgjmop11}. Similarly to previous $\alpha_s$ determinations
from $\tau$ decays \cite{bck08,ddhmz08,bj08,my08}, the FOPT result turns out
to be lower than the CIPT value, though due to the additional DV parameters
the errors turn out larger and the difference is less significant.

\begin{figure}[!ht]
\begin{center}
\includegraphics[width=0.9\columnwidth,angle=0]
{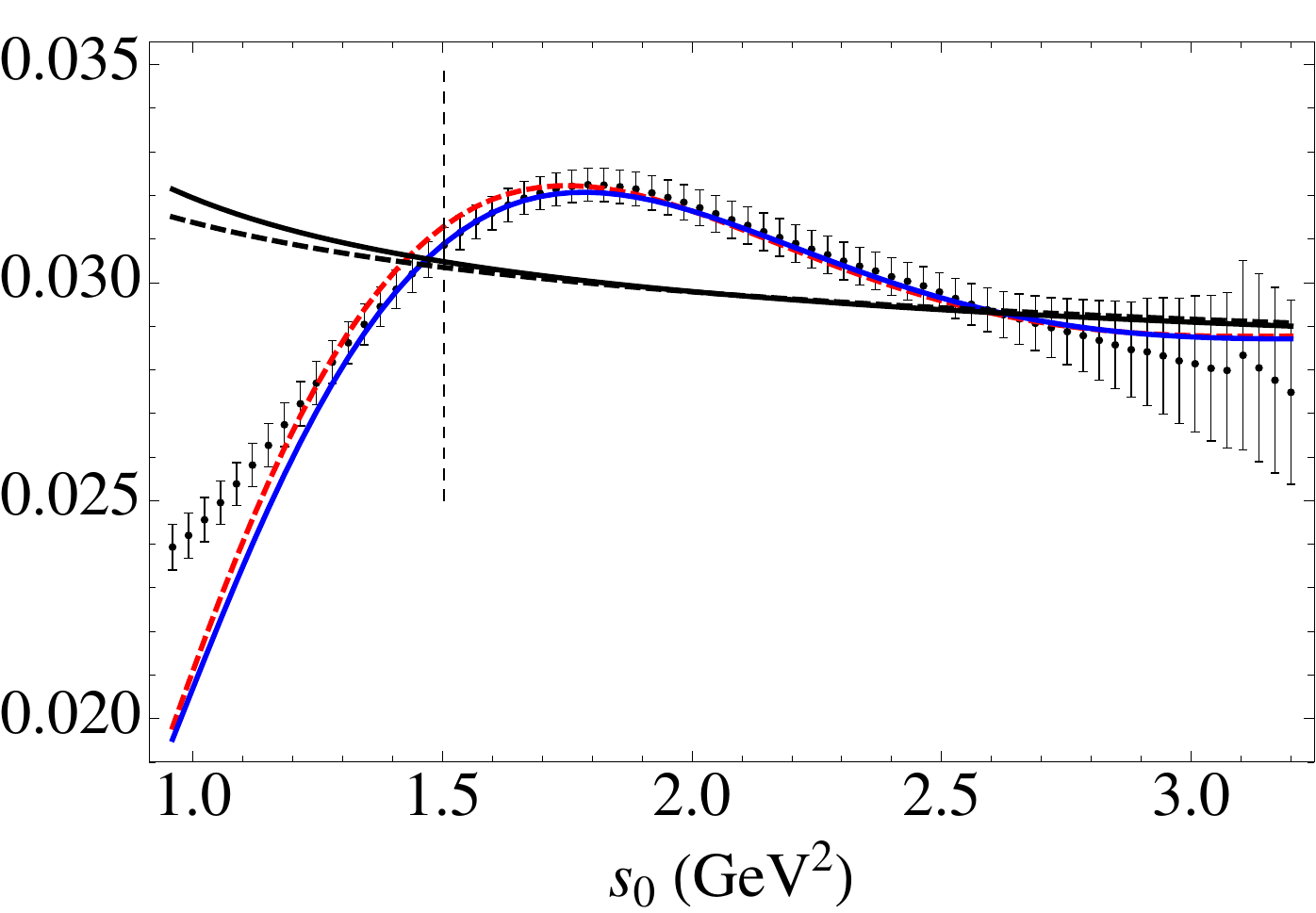}
\vspace{-0.2cm}
\caption{Comparison of experimental and theoretical $w(s)=1$ $A$ moments. FOPT
fits are shown in blue (solid) and CIPT fits in red (dashed). The (much flatter)
black curves display only the OPE parts of the FOPT (solid) and CIPT (dashed)
fit results.\label{fig3A}}
\end{center}
\end{figure}

The next level of sophistication is including the axial channel in the fit with
$w(s)=1$. This introduces the axial DV parameters as additional fit parameters.
It is found, however, that a fit employing the full set of $s_0$, corresponding
to all right bin endpoints above $s_{\rm min}$, is no longer possible. A stable
fit is still obtained, though, if only every third $s_0$ is included. The
comparison of the vector moments in this case looks very much like figure~1.
In figure~2, the corresponding comparison is given for the $A$ moments, with
the notation being the same as for figure~1. Again, a fit without DVs would
not provide an acceptable description of the data. The resulting values for
$\alpha_s$ are found to be
\begin{eqnarray}
\label{asw1}
\alpha_s(M_\tau)\!\!\!&=&\!\!\! 0.308\pm 0.016\pm 0.009\quad\mbox{(FOPT)}\,,\\
\alpha_s(M_\tau)\!\!\!&=&\!\!\! 0.325\pm 0.022\pm 0.011\quad\mbox{(CIPT)}\,,
\end{eqnarray}
very similar to the vector only case, but somewhat less stable under variation
of $s_{\rm min}$. A full account of all fit parameters is found in table~3 of
\cite{bcgjmop11}.

Regarding fits including several weight functions, the additional weights
$w(x)=1-x^2$ with $x\equiv s/s_0$ and the kinematical weight
$w(x)=(1-x)^2(1+2x)$ have been investigated by us in detail. The first one has
a single pinch suppression and includes the $D=6$ OPE term while the second is
doubly pinched and involves condensates up to $D=8$. As soon as a second weight
is added to the fit, even reducing the number of employed $s_0$ only leads to
very unstable fits, due to the strong correlations. Thus, for these fits we
have decided to follow a different route. Here the cross-correlations between
different weights are dropped in the fit, but later again included by a linear
fluctuation analysis. This again yields results fully compatible with the ones
given above, though without a reduction in the error of $\alpha_s$.

\begin{figure}[!th]
\begin{center}
\includegraphics[width=0.9\columnwidth,angle=0]
{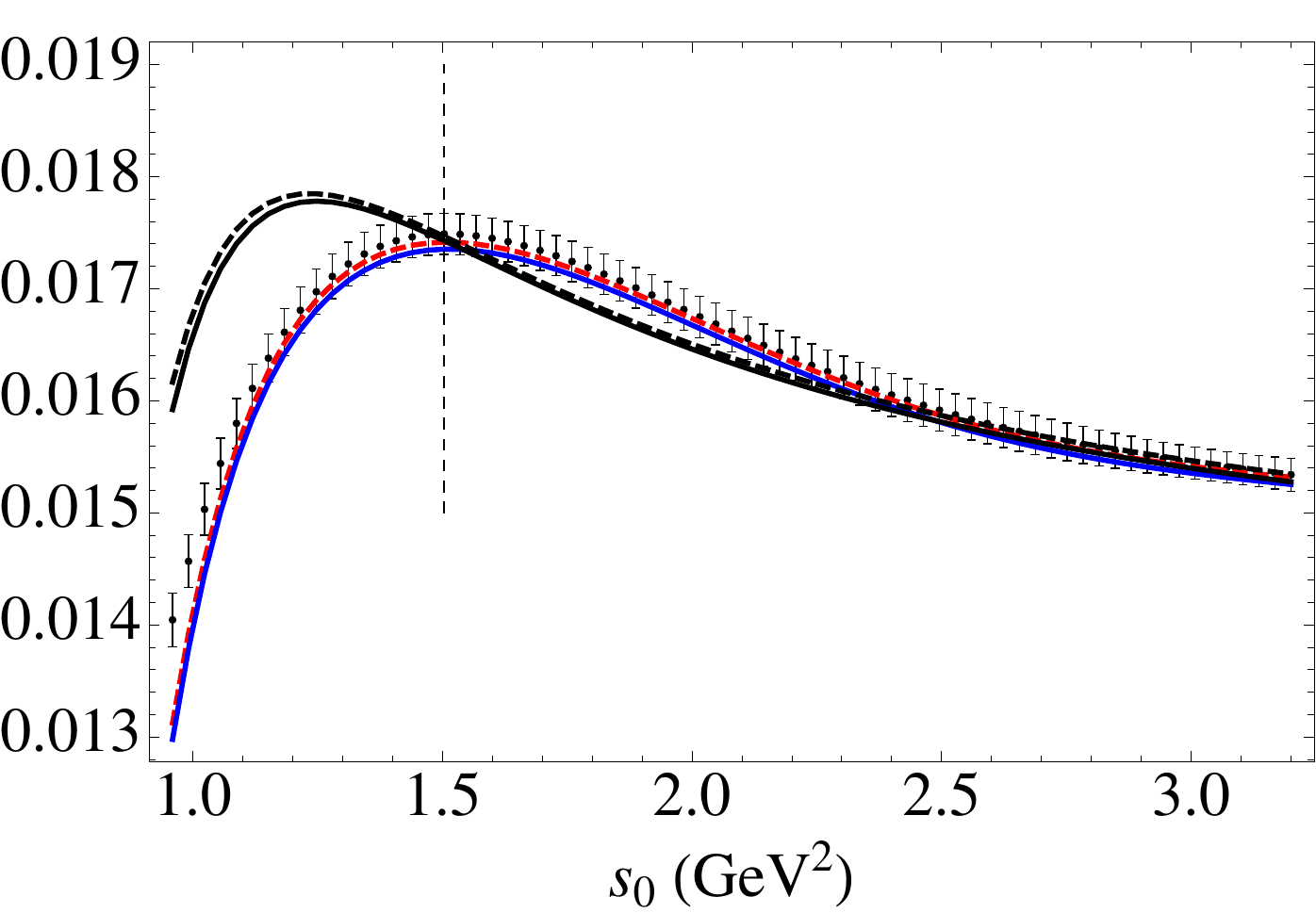}
\vspace{-0.2cm}
\caption{Comparison of experimental and theoretical $w(x)=(1-x)^2(1+2x)$ $V$
moments. FOPT fits are shown in blue (solid) and CIPT fits in red (dashed).
The black curves display only the OPE parts of the FOPT (solid) and CIPT
(dashed) fit results.\label{fig4V}}
\end{center}
\end{figure}

As one example figure~3 displays the moment comparison for the kinematical
weight. Due to the double pinch suppression, the pure OPE fares much better
even down to lower energies. Still, below roughly $s_0\approx 2.5\,{\rm GeV}^2$
differences to the full fit including DVs are clearly visible. Furthermore,
the inclusion of the DVs also influences the values of the fitted condensate
parameters. This is exemplified by a more detailed discussion of the $D=6$
condensates and violations of the vacuum-saturation approximation which can be
found in section VI.B of ref.~\cite{bcgjmop11}.

\section{Conclusions}
\label{sect4}
Based on ref.~\cite{bcgjmop11}, a brief summary of the determination of
$\alpha_s$ from hadronic $\tau$ decays including effects due to duality
violations in a consistent fashion has been presented above. The central
result will be taken from eqs.~(\ref{asw0}) and (9), since as far as OPE
contribution is concerned, the weight $w=1$ is most clean. Evolving this
result to the $Z$ boson mass, one obtains:
\begin{eqnarray}
\label{asw2}
\alpha_s(M_Z)\!\!\!&=&\!\!\! 0.1169\pm 0.0025 \quad\mbox{(FOPT)}\,,\\
\alpha_s(M_Z)\!\!\!&=&\!\!\! 0.1187\pm 0.0032 \quad\mbox{(CIPT)}\,.
\end{eqnarray}
Our results are somewhat lower than the most frequently quoted previous
$\alpha_s$ values from $\tau$ decays, but on the other hand the uncertainty
is increased in view of the additional degrees of freedom in the fit through
the DVs.

The analysis presented in \cite{bcgjmop11} and this contribution was based on
the original OPAL data \cite{opal99}, in order to clearly be able to compare to
this analysis and single out the effect of including DVs. The corresponding
results by OPAL for $\alpha_s(M_Z)$ read $0.1191(16)$ and $0.1219(20)$ for FOPT
and CIPT respectively.

As an obvious next step, the analysis should be repeated with an updated data
set which includes up-to-date $\tau$ branching fractions and values of other
inputs. Such an analysis is currently under way. Then, for comparison, an
analysis of the eventually updated ALEPH data would be most helpful, especially
as the errors from this data set are expected to be smaller. Finally, in the
long run, it is to be hoped that also the B-factories BaBar and Belle will at
some point provide the complete $V$ and $A$ spectral functions, since with the
available statistics, substantial improvements over the LEP experiments are to
be envisaged.

\vspace{3ex}
\noindent {\bf Acknowledgments}

{We would like to thank Sven~Menke for significant help with understanding
the OPAL spectral-function data. DB, MJ and SP are supported by
CICYTFEDER-FPA2008-01430, SGR2005-00916 and the Spanish Consolider-Ingenio 2010
Program CPAN (CSD2007-00042). SP is also supported by a fellowship from the
Programa de Movilidad PR2010-0284. OC is supported in part by MICINN (Spain)
under Grant FPA2007-60323, the Spanish Consolider-Ingenio 2010 Program CPAN
(CSD2007-00042) and by the DFG cluster of excellence ``Origin and Structure of
the Universe.'' MG and JO are supported in part by the US Department of Energy,
and KM is supported by a grant from the Natural Sciences and Engineering
Research Council of Canada.}



%% References

\nocite{*}
%\bibliographystyle{elsarticle-num}
%\bibliography{jamin_phi2psi_2011}


\end{document}